\documentclass[fleqn,usenatbib]{mnras}

\usepackage{newtxtext,newtxmath}
\usepackage[T1]{fontenc}

\DeclareRobustCommand{\VAN}[3]{#2}
\let\VANthebibliography\thebibliography
\def\thebibliography{\DeclareRobustCommand{\VAN}[3]{##3}\VANthebibliography}

\usepackage{graphicx}	% Including figure files
\usepackage{amsmath}	% Advanced maths commands
\usepackage{xspace}
\usepackage{booktabs}
\usepackage{siunitx}
\usepackage{caption}
\usepackage{longtable}
\usepackage{pdflscape}
\newcommand{\Teff}{\ensuremath{T_{\rm eff}}\xspace}
\newcommand{\numax}{\ensuremath{\nu_{\rm max}}\xspace}
\newcommand{\dnu}{\ensuremath{\Delta\nu}\xspace}
\newcommand{\bprp}{$G_{\rm bp}-G_{\rm rp}$\xspace}
\newcommand{\eps}{\ensuremath{\varepsilon_{\rm p}}\xspace}

\newcommand{\muHz}{\ensuremath{\mu}Hz\xspace}

\newcommand{\pbjam}{\texttt{PBjam}\xspace}

\newcommand{\kepler}{\textit{Kepler}\xspace}
\newcommand{\corot}{\textit{CoRoT}\xspace}
\newcommand{\tess}{\textit{TESS}\xspace}
\newcommand{\plato}{\textit{PLATO}\xspace}
\newcommand{\ktwo}{\textit{K2}\xspace}
\newcommand{\gaia}{\textit{Gaia}\xspace}

\title[Peakbagging KEYSTONE sample with PBjam]{Peakbagging the K2 KEYSTONE sample with PBjam: characterising the individual mode frequencies in solar-like oscillators}

\author[G. T. Hookway et al.]
{George T. Hookway $^{\href{https://orcid.org/0009-0002-8134-4026}{\includegraphics[scale=0.3]{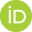}}}$,$^{1}$\thanks{E-mail: gth025@student.bham.ac.uk}
Martin B. Nielsen $^{\href{https://orcid.org/0000-0001-9169-2599}{\includegraphics[scale=0.3]{Figures/orcid.png}}}$,$^{1}$
Guy R. Davies $^{\href{https://orcid.org/0000-0002-4290-7351}{\includegraphics[scale=0.3]{Figures/orcid.png}}}$,$^{1}$
Mikkel N. Lund $^{\href{https://orcid.org/0000-0001-9214-5642}{\includegraphics[scale=0.3]{Figures/orcid.png}}}$,$^{2}$\newauthor
Rafael A. García $^{\href{https://orcid.org/0000-0002-8854-3776}{\includegraphics[scale=0.3]{Figures/orcid.png}}}$,$^{3}$
Savita Mathur $^{\href{https://orcid.org/0000-0002-0129-0316}{\includegraphics[scale=0.3]{Figures/orcid.png}}}$,$^{4,5}$
Victor See $^{\href{https://orcid.org/0000-0001-5986-3423}{\includegraphics[scale=0.3]{Figures/orcid.png}}}$$^{1}$
and Amalie Stokholm $^{\href{https://orcid.org/0000-0002-5496-365X}{\includegraphics[scale=0.3]{Figures/orcid.png}}}$$^{1,2}$
\\
$^{1}$School of Physics and Astronomy, University of Birmingham, Edgbaston, Birmingham, B15 2TT, UK\\
$^{2}$Stellar Astrophysics Centre, Department of Physics and Astronomy, Aarhus University, Ny Munkegade 120, DK-8000 Aarhus C, DK\\
$^{3}$Université Paris-Saclay, Université Paris Cité, CEA, CNRS, AIM, 91191, Gif-sur-Yvette, France\\
$^{4}$Instituto de Astrofíısica de Canarias (IAC), E-38205 La Laguna, Tenerife, Spain\\
$^{5}$Universidad de La Laguna (ULL), Departamento de Astrofíısica, E-38206 La Laguna, Tenerife, Spain
}

\date{Accepted 2025 October 21. Received 2025 September 19; in original form 2025 June 25}

\pubyear{\the\year{}}

\begin{document}
\label{firstpage}
\pagerange{\pageref{firstpage}--\pageref{lastpage}}
\maketitle

\begin{abstract}
The pattern of individual mode frequencies in solar-like oscillators provides valuable insight into their properties and interior structures. The identification and characterisation of these modes requires high signal-to-noise and frequency resolution. The KEYSTONE project unlocks the asteroseismic potential of the \ktwo mission by providing individually reduced, high-quality time series data, global asteroseismic parameters, and spectroscopic analysis for 173 solar-like oscillators. In this work, we build on the KEYSTONE project and present the first analysis of the pattern of individual modes in the oscillation spectra for the \ktwo KEYSTONE stars. We perform a robust identification and characterisation of the modes through peakbagging methods in the open-source analysis tool \pbjam. We present over 6000 mode frequencies, widths, and heights for 168 stars in the sample, covering the HR diagram from FGK dwarfs to sub-giants and the lower red giant branch, providing a significant increase in the number of individual mode frequency detections for main sequence and sub-giant oscillators. This study also presents sample-wide trends of oscillation patterns as a function of the fundamental stellar properties, and improves the precision of the global asteroseismic parameters. These measurements are part of the legacy of the \ktwo mission, and can be used to perform detailed modelling to improve the precision of fundamental properties of these stars. The results of this analysis provides evidence for the validity of using \pbjam to identify and characterise the modes resulting from the observations of the future \plato mission.
\end{abstract}

\begin{keywords}
physical data and processes: asteroseismology - stars: oscillations - stars: solar-type
\end{keywords}

%%%%%%%%%%%%%%%%%%%%%%%%%%%%%%%%%%%%%%%%%%%%%%%%%%

%%%%%%%%%%%%%%%%% BODY OF PAPER %%%%%%%%%%%%%%%%%%

%\tableofcontents

\section{Introduction} \label{sec: Introduction}

Over the last few decades, missions such as \corot~\citep{Baglin_2006,Auvergne_2009}, \kepler~\citep{Gilliland_2010}, \ktwo~\citep{Howell_2014}, and \tess~\citep{Ricker_2014} have measured the brightness of millions of stars with high photometric precision and thus provided valuable insight into the internal oscillations of thousands of solar-like oscillators. The high-quality data from these missions have helped improve the understanding of diverse areas of stellar physics (e.g. see reviews by~\citealt{Chaplin_2013,dimauro2017reviewasteroseismology,Garc_a_2019}), such as stellar evolution~\citep{Bedding_2011,Silva_Aguirre_2012}, inference of stellar properties~\citep{Rendle_2019,BASTA_2021,Lyttle_2021}, exoplanet host star characterisation (e.g.~\citealt{Gilliland_2010-2};~\citealt*{Metcalfe_2015}), internal rotation~\citep[e.g.][]{Beck_2012,Deheuvels_2012,Davies_2015-2,Di_Mauro_2016}, rotation evolution \citep[e.g.][]{Hall_2021}, and internal magnetic fields~\citep[e.g.][]{Fuller_2015, Stello_2016, Hatt_2024}.

The \ktwo asteroseismic KEYSTONE project~\citep[][see also~\citealt{Chaplin_2015,Lund_2016}]{Lund_2024} aims to build upon these previous works and strengthen our understanding of stellar physics by providing a sample of stars with global asteroseismic parameters. The sample consists of 173 stars observed by the \ktwo mission from campaigns 6-19 in short cadence mode ($\Delta t=58.89$s), which allows for asteroseismic analyses of main sequence stars. These stars have \kepler photometric magnitudes in the range 6-13, which places them at the brighter end of the \kepler magnitude range. The sample consists primarily of main sequence FGK stars, although the sample also contains sub-giants and even some low-luminosity red giants.

~\citet{Lund_2024} have already performed spectroscopic and asteroseismic analyses for all the stars in the catalogue, providing estimates for the global asteroseismic parameters, namely the frequency of maximum power, \numax, and the large frequency separation, \dnu. These parameters are closely related to surface gravity and mean density, respectively~\citep{Ulrich_1986,Brown_1991,Kjeldsen_1995}. When combined with atmospheric effective temperature, \Teff, they can provide strong constraints on radius and mass, leading to tight constraints on age when used in conjunction with stellar models~\citep{Aerts_2015,Garc_a_2019}.

The individual mode frequencies can be used to probe different parts of the stellar interior, meaning the mode properties provide more information about the stellar structure than the global asteroseismic parameters. This allows for improved constraints on the fundamental stellar properties and structure, such as composition, mass and age \citep[e.g.][]{Mathur_2012,Lebreton_2014,Metcalfe_2014}. One way to acquire these individual frequencies is through ``peakbagging'' methods, where the oscillation peaks in a power spectrum are characterised and identified using their radial order, $n$, angular degree, $\ell$, and azimuthal order $m$ (see~\citealt{Appourchaux_2012,Davies_2015,Lund_2017}). In addition, previous ensemble peakbagging studies have provided individual mode properties for other large stellar samples, including giant stars~\citep[e.g.][]{Vrard_2018,Li_2020}.

In this work, we will utilise \pbjam~\citep{Nielsen_2021,Nielsen_2025}, which allows for analysis of mode properties without the need for manual mode identification. The input light curve and oscillation envelope model provide the likelihood of the frequency location of each individual mode. This is combined with a large prior sample of thousands of stars to create a posterior probability density for the frequencies of the individual modes. Models are fit to the data for a specified number of radial orders, with \pbjam outputting a series of $\ell=0,1,2$ individual mode frequencies.

In this study, we aim to use \pbjam to perform the first identification of the individual $\ell=0,1,2$ modes for the stars in the KEYSTONE sample. The identified modes can then be peakbagged in order to extract the properties of the individual mode frequencies for the 173 stars in the sample. These mode properties can then be used with stellar models in future studies, allowing us to build the foundations for improved precision on the fundamental characteristics of these stars as well as stars of similar evolution and masses.

The remainder of this paper is orgainsed as follows: in Sec.~\ref{sec: Input Data} we discuss the selection of stellar parameters to be used in the mode identification of \pbjam, along with the light curve data and how we dealt with instrumental variability. Sec.~\ref{sec: PBjam} discusses the process of mode identification and peakbagging with \pbjam, along with the process of ensuring the quality of the results. Sec.~\ref{sec: Results} provides the results for the data set as a whole, along with individual results for an example set of stars. Finally, in Sec.~\ref{sec: Discussion} we summarise our conclusions.

\section{Input Data} \label{sec: Input Data}

In this study, we apply peakbagging techniques to the 173 stars in the KEYSTONE sample. The peakbagging algorithm of \pbjam (see Sec.~\ref{sec: PBjam}) requires stellar parameters \numax, \dnu, \Teff, and \bprp to be used to select a suitable prior sample, as well as the power density spectrum (PDS) for each of the stars. In this section, we outline the process of obtaining the stellar parameters, along with the treatment of the light curve data that resulted in the PDS, provided by~\citet{Lund_2024}.

\begin{figure}
    \centering
    \includegraphics[width=0.96\linewidth]{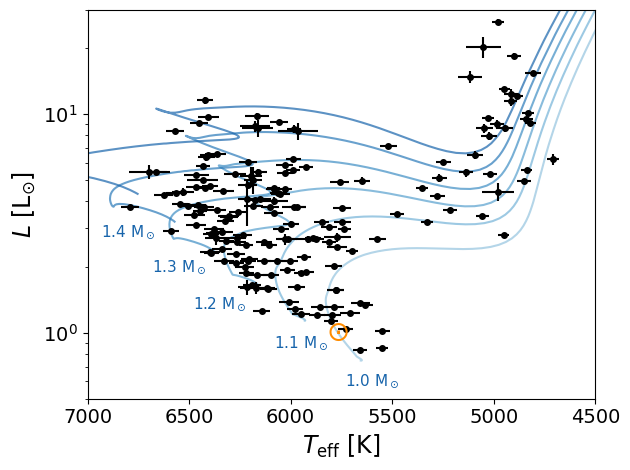}
    \caption{HR diagram of the 173 stars in the KEYSTONE sample, which are analyzed in this work. The evolutionary tracks (blue) for masses 1.0-1.5 M$_\odot$ are shown for [Fe/H]=0. The position of the Sun is shown for reference.}
    \label{fig: Sample}
\end{figure}

We used \Teff estimates from~\citet{Lund_2024} that were calculated with the Infrared Flux Method (IRFM;~\citealt{Casagrande_2010}) and also from the Stellar Parameter Classification pipeline (SPC;~\citealt{Buchave_2012}). Both these pipelines produce \Teff values that are in good agreement, so we used IRFM results for the majority of stars and SPC results for 3 stars where IRFM results were not available. For the asteroseismic parameters \numax and \dnu, we used the results from the coefficient of variation (CV;~\citealt*{Bell_2018}) method and the SYD pipeline~\citep{Huber_2009}. Again, both these pipelines are in good agreement on \numax and \dnu, so we used results from the CV method for the majority of stars and SYD results for the remaining 3 stars where CV results were unavailable. For each star, we used \bprp values from \gaia DR3~\citep{Gaia_2016, Gaia_2023}, with an uncertainty of $0.1$ dex applied to all the stars. The Hertzsprung-Russell (HR) diagram in Fig.~\ref{fig: Sample} shows the sample used in this analysis. % Add a separate HR diagram with no colourings, with numax vs Teff?

\begin{figure*}
    \centering
    \includegraphics[width=0.96\linewidth]{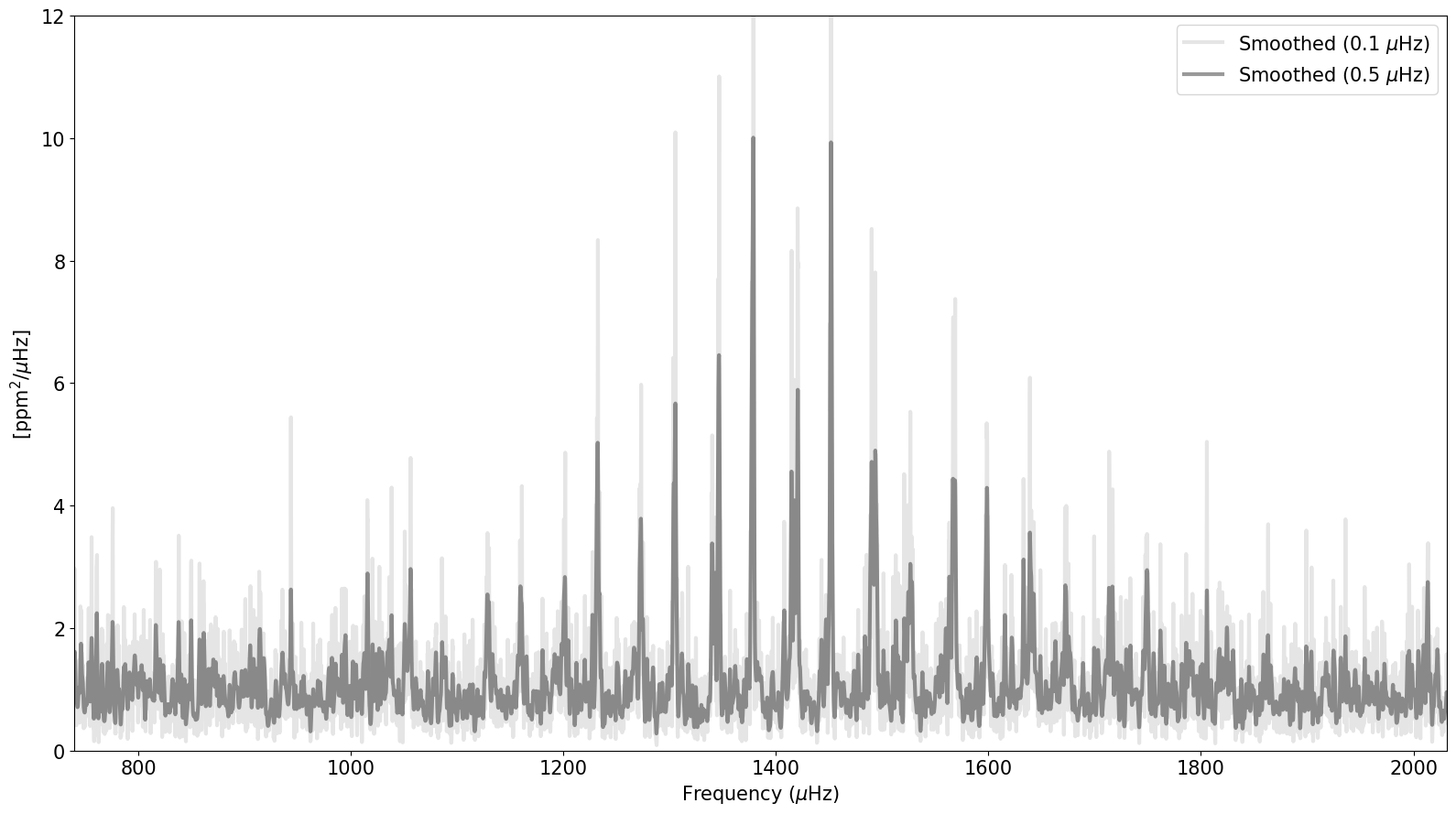}
    \caption{Power density spectrum for EPIC 241011563, showing the increased level of power from the p-mode oscillation envelope. For clarity the spectrum has been smoothed with a Gaussian kernel with widths of $0.1\mu$Hz (light grey) and $0.5\mu$Hz (dark grey).}
    \label{fig: PDS}
\end{figure*}

The light curve data from the \textit{K2} observations for each of our stars were processed using the K2P$^2$ pipeline~\citep{Lund_2015,Lund_2024}. This pipeline reduces the effects from systematic variability, including the roll of the spacecraft and the removal of instrumental and potential planetary signals via the Kepler Asteroseismic Science Operations Center (KASOC) filter~\citep{Lund_2015}. The pipeline provides the PDS (see~\citealt{Handberg_2014}) for this sample of stars (see Fig.~\ref{fig: PDS} for an example PDS), to which we applied our peakbagging algorithm.

The observations for 30 stars in the sample took place over two observing campaigns. The time between these observing campaigns has the effect of leaving gaps in the time series. When the power spectra for these stars are created, artifacts appear in the data due to the discontinuity in the data. However, the large gaps in time between observations translates to minimal spectral leakage, meaning the artifacts would not have a significant impact on the analysis.

\begin{figure}
    \centering
    \includegraphics[width=0.96\linewidth]{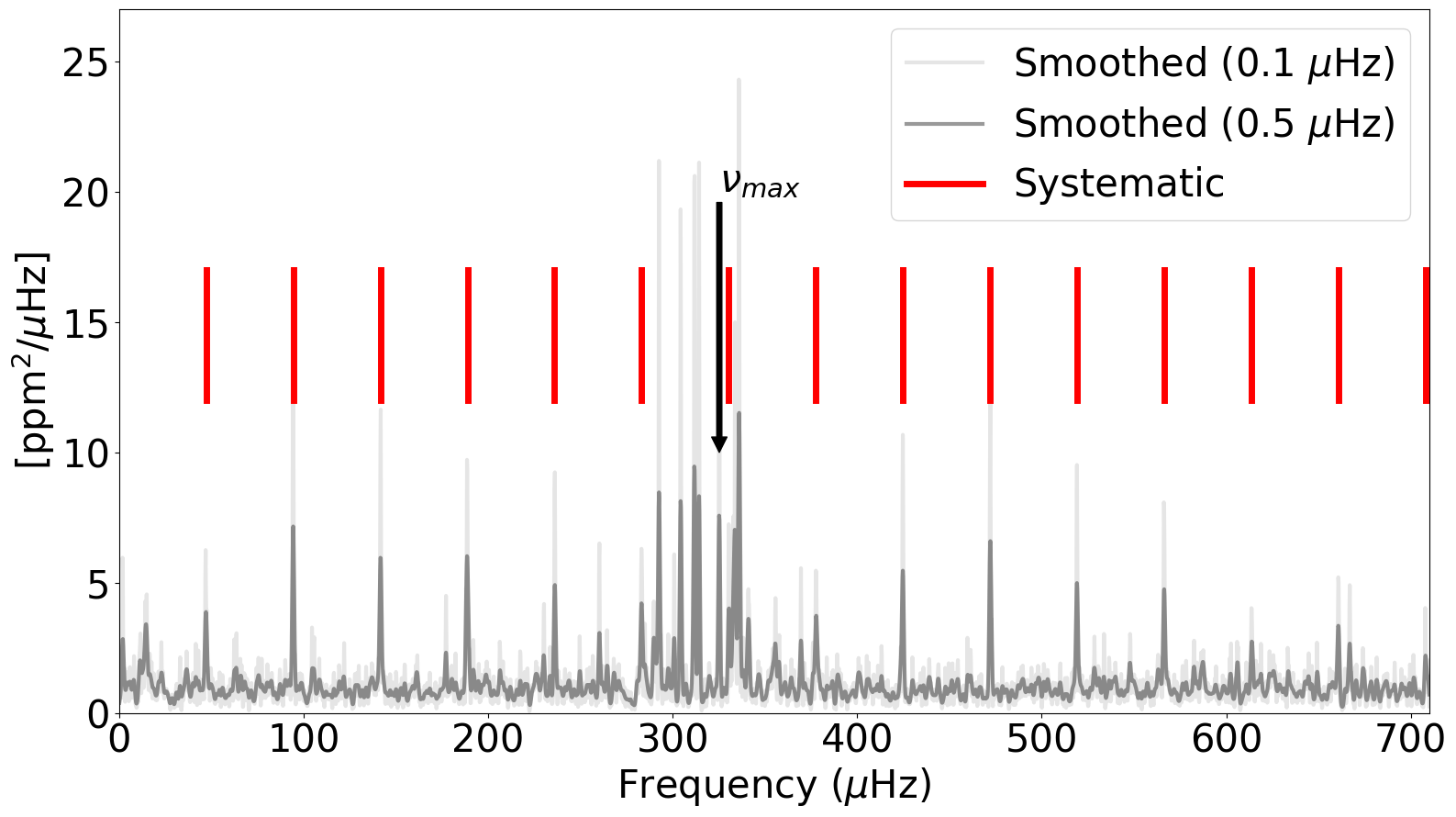}
    \caption{Power spectrum of EPIC 212586030, a star with prominent overtones of instrumental variability caused by the roll of the spacecraft. The black arrow shows the location of \numax, and the solid red lines show the location of the instrumental variability overtones, every $\sim47.2\mu$Hz.}
    \label{fig: Systematics}
\end{figure}

The systematic variability caused by the roll of the spacecraft causes power to appear in the PDS at multiples of ${\sim}47.2$ \muHz \citep{Lund_2015}. To prevent these from being mistaken for actual mode frequencies, we omitted frequency bins within $1$ \muHz of the instrumental peak for spectra where these systematics showed significant power based on a manual inspection. Above $700$ \muHz, the power of these systematics has a low enough impact on the spectrum to not require removal. Fig.~\ref{fig: Systematics} shows the power spectrum of EPIC 212586030, which has prominent power from these systematic signals in the PDS, resulting in the removal of two $\ell=1$ modes and one $\ell=2$ mode. In cases where this data removal encompassed a mode, the modes were removed from the spectrum and the analysis. Since this instrumental variability is only prominent at frequencies below ${\sim}700$~\muHz, this only affected 6 of the sub- and red giant branch stars, in which case only 1 or 2 modes were removed from the spectra.

Other common systematics that appear in the spectra occur at ${\sim}1650$ and ${\sim}2100$~\muHz. We found 10 stars where the systematic lined up with, or was close enough to, a mode to cause confusion for the automated mode identification. In these cases the affected data points were removed, with a range determined by a manual inspection of the spectrum. This results in a fraction of a radial order being removed from the spectrum of these stars, leading to the removal of between 1-3 mode frequencies.

\section{Peakbagging} \label{sec: PBjam}

Peakbagging is the process of fitting a model to the power spectrum of a stellar light curve with the aim of estimating the individual mode frequencies for a star~\citep{Appourchaux_2003}. The peakbagging tool \pbjam~\citep{Nielsen_2021, Nielsen_2025} was used to estimate the mode frequencies for the stars in our sample. \pbjam performs peakbagging using a Bayesian framework, with it sampling posterior distributions following Bayes's theorem,
\begin{equation}
\ln{P(\boldsymbol{\theta}|D)}\propto \ln{\mathcal{L}\left(D|\boldsymbol{\theta}\right)} + \ln{P(\boldsymbol{\theta})}.
\label{eq: Bayes}
\end{equation}
Here, $\ln{P(\boldsymbol{\theta})}$ is the prior probability of drawing model parameters $\boldsymbol{\theta}$, and $\ln{\mathcal{L}\left(D|\boldsymbol{\theta}\right)}$ is the likelihood of the data being observed given these model parameters. The model of the oscillations in the spectrum is treated as the sum of Lorentzian profiles~\citep*{Anderson_1990},
\begin{equation}
M(\nu) = \sum\limits_{n}\sum\limits_{\ell}\sum\limits_{m=-\ell}^{\ell}\frac{H_{n \ell m} }{1 + \frac{4}{\Gamma_{n\ell}^2}(\nu-\nu_{n \ell m})^2},
\label{eq: model}
\end{equation}
shown here in a general form. The first sum in the equation is over the number of p and g modes in the model which will vary depending on the angular degree of the oscillations being included. The second sum is over these particular angular degrees in a given model, and the third sum is over the $m=2\ell+1$ azimuthal orders for each angular degree. $H_{n \ell m}$ and $\Gamma_{n\ell}$ represent the height and  mode width of the modes, respectively. \pbjam performs this Bayesian inference based peakbagging in two main stages, the mode identification followed by the final peakbagging stage.

\subsection{Mode Identification} \label{sec: Mode ID}

Mode identification is the process of finding peaks in the power spectrum of a star that we believe to be the modes of oscillation. These peaks are assigned the identifiers: angular degree, $\ell$, radial order, $n$, and azimuthal order, $m$. The first step in mode identification with \pbjam involves creating a model for the background noise and the $\ell=2,0$ modes. The background noise model consists of four terms: the shot noise and three Harvey-like profiles~\citep{Harvey_1985}. The background model includes contributing terms from instrumental variability, rotation and magnetic activity, as well as convection on the stellar surface. The model for the $\ell=2,0$ modes consists of the sum of Lorentzian profiles from equation~\ref{eq: model}. For the $\ell=2,0$ model, the radial order sum is over the input number of radial orders in the model, $N_\mathrm{p}$, centred around \numax. We varied $N_\mathrm{p}$ for each star based on the S/N of the power spectra from a manual inspection, with it ranging from $N_\mathrm{p}=5$ to $N_\mathrm{p}=20$ for the lower and higher S/N cases, respectively. We chose to be optimistic with the number of orders included to not miss any modes. The radial order sum in equation~\ref{eq: model} takes the limits $n_\mathrm{max}-N_\mathrm{p}/2$ and $n_\mathrm{max}+N_\mathrm{p}/2$ for the $\ell=2,0$ model, with $n_\mathrm{max}$ being the radial order at \numax, defined by $n_{\mathrm{max}}=\numax/\dnu - \eps$, where \eps is the phase shift. The sum over the azimuthal orders for the $\ell=2,0$ model results in  a single $\ell=0$ and a multiplet of five evenly split $\ell=2$ modes for each $\ell=2,0$ pair, allowing for the rotational splitting of the $\ell=2$ modes~\citep*{Gizon_2003,Ballot_2006}. The multiplets are assumed to experience the same splitting, regardless of radial order. For a given $\ell=2,0$ pair, the $\ell=2$ multiplets are separated from the $\ell=0$ modes by a small frequency separation, $\delta\nu_{02}$. This results in the frequency of the of the $\ell=2$ modes being defined by $\nu_{n_{\mathrm p}-1,2} = \nu_{n_{\mathrm p},0}-\delta\nu_{02}$.

The $\ell=0$ mode frequencies, $\nu_{n_\mathrm{p},l=0}$, from equation~\ref{eq: model} are parameterised following an asymptotic relation, 

\begin{equation}
    \nu_{n_\mathrm{p}, 0} =\dnu\left(n_\mathrm{p}+\eps+\frac{\alpha_\mathrm{p}}{2}\left(n_\mathrm{p}-n_{\mathrm{max}}\right)^2\right),
\label{eq: asymptotic}
\end{equation}
where $\alpha_\mathrm{p}$ determines the scale of the second order variation of the mode frequencies with radial order. The mode widths are parameterised following the final part of equation~\ref{eq: model}, where the mode widths for a given star are treated as a constant $\Gamma$ for the central few modes around \numax. The mode heights of the Lorentzian profiles, $H_{n \ell m}$, are parameterised as a Gaussian envelope centred around \numax. The heights of the modes in the multiplets caused by the splitting are determined by the azimuthal order of the mode and the angle of inclination between the line of sight of the observer and the rotational axis of the star~\citep{Nielsen_2025}.

The second step of the mode identification with \pbjam involves applying a model of the $\ell=1$ modes. The previous $\ell=2,0$ and background models leave a residual spectrum which includes the $\ell=1$ modes. The model for the $\ell=1$ p modes is once again the sum of Lorentzian profiles, shown in equation~\ref{eq: model}. For the $\ell=1$ modes, the model used is much the same, though now the sum over the radial orders is over $N_\mathrm{p}+N_\mathrm{g}$, where $N_\mathrm{g}$ is the number of g mode radial orders included in the model. The inclusion of the g modes accounts for the coupling between the p and g modes. As stars evolve off the main sequence and become sub- and red giant stars, the $\ell=1$ p and g-modes can become mixed. \pbjam can account for this, and provides a more complex model for the $\ell=1$ modes for these stars, allowing for the fitting of mixed modes in the spectrum.

\begin{figure}
    \centering
    \includegraphics[width=0.96\linewidth]{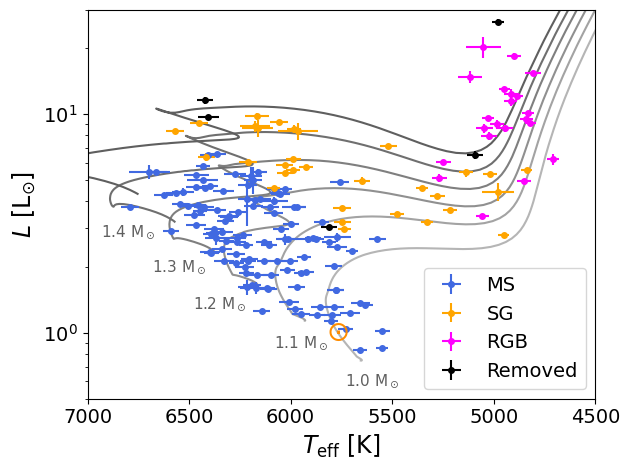}
    \caption{HR diagram of the 173 stars in the KEYSTONE sample. The colours denote the different fitting models used in \pbjam, with main sequence (MS) in blue, sub-giant (SG) in orange, and red giant branch (RGB) in magenta. The black points denote the five stars that were removed from the sample (see Sec.~\ref{sec: Global Results}). The 1.0-1.5 M$_\odot$ evolutionary tracks (grey) are shown for [Fe/H] = 0. The position of the Sun is shown for reference.}
    \label{fig: Star Types}
\end{figure}

\pbjam has three different models for fitting the dipole modes, main sequence (MS), sub-giant (SG) and red giant branch (RGB)~\citep{Nielsen_2025}. The names for these models do not necessarily represent the physics of the star, but are instead based on the visibility of mixed modes in the spectrum. The MS model is the basic model where the modes behave like pure p modes with no signs of g-mode coupling, so in this model $N_g=0$. In this case the $\ell=1$ modes behave in the same asymptotic manor as the $\ell=2,0$ pair. The modes are separated from the $\ell=0$ modes by a small frequency separation, $\delta\nu_{01}$, which defines the $\ell=1$ frequency as,
\begin{equation}
    \nu_{n_{\mathrm p},1} = \nu_{n_{\mathrm p},0}+\delta\nu_{01}.
    \label{eq: l=1 p-modes}
\end{equation}
The SG model is used for stars where the mixed modes are now visible, though typically with fewer than two g modes per radial order. The dipole g-modes are separated by the period spacing, $\Delta\Pi_1$, and are defined by,
\begin{equation}
    \nu_{n_{\mathrm g},1} = [\Delta\Pi_1(n_{\mathrm g}+\varepsilon_{\mathrm g})]^{-1},
    \label{eq: l=1 g-modes}
\end{equation}
where $\varepsilon_{\mathrm g}$ is the phase offset of the g-modes. The mixed mode frequencies are now modelled by a non-asymptotic description~\citep{Deheuvels_2010,Ong_2020}. The solution to a general Hermitian eigenvalue equation provides the weighting between the pure p and g modes that describes the strength of the mode coupling. For more than two g modes per radial order, the RGB model is used. This model best describes the p- and g- mode cavities separately and uses a uniform coupling model to find the solutions for the mixed mode frequencies. The RGB model also allows for asymmetric splitting of the mixed $\ell=1$ modes, which allows for significant differences in the envelope and core rotation rates. The value for $N_g$ ranges from 1-10 for the SG model and can be several dozen for the RGB model~\citep{Nielsen_2025}. The initial choice of $\ell=1$ model used for each star was based on an inspection of the HR diagram of the sample (see Fig.~\ref{fig: Star Types} for the choice of model for each star). The models chosen for stars that were on the MS-SG or SG-RGB boundary were decided based on an inspection of which model worked best for that particular star.

The g modes propagate horizontally and so do not couple with the radial p modes~\citep*{Aerts_2010}, so there is no need for a more complex $\ell=0$ model. The g modes do couple with modes of $\ell>0$, though since the evanescent region increases with $\ell$, the coupling between the p and g mode cavities for the $\ell=2$ is weak and mixed modes only rarely appear in the power spectra. The majority of the modes are therefore almost completely p-like $\ell=2$ modes and follow the asymptotic relation for p modes in equation~\ref{eq: asymptotic} throughout the evolution of the star. \pbjam only accounts for mixed $\ell=1$ modes, while no mixing model is available for $\ell=2$ modes due to it becoming computationally expensive. 

\subsubsection{Bayesian Inference}

To estimate the probability density of each of the parameters in the mode identification stage and, by extension, the mode frequencies, \pbjam uses the \texttt{Dynesty} package for Python~\citep{Speagle_2020}. \texttt{Dynesty} utilises nested sampling, to account for multimodal posterior solutions~\citep{Skilling_2004,Feroz_2013} which may be caused by the combination of different Lorentzian profiles. The posterior probability distributions are estimated by drawing samples from the model parameter space within the boundaries set-out by the prior and are weighted based on the likelihood function.

The dataset of prior stars consists of thousands of stars from \emph{Kepler} and \textit{TESS}, along with stellar model grid samples (see~\citealt{Nielsen_2025}). For the fitting of the $\ell=2,0$ modes, a prior sample was selected based on the nearest sample of 50 stars to the input parameters \numax, \dnu, and \Teff from the larger prior dataset, while the prior sample selected for the $\ell=1$ modes consisted of 100 stars. The number of prior stars selected was a balance of speed and accuracy. Many of the parameters in the model are highly correlated, with them depending on the same fundamental stellar properties. This allows for the method of dimensionality reduction described by~\citet{Nielsen_2023} to be used in the construction of the prior. The prior probability densities are constructed from one-dimensional kernel-density estimates in the latent parameter space~\citep{Nielsen_2025}.

The global asteroseismic parameters \dnu and \numax are strong indicators of the form of the power spectrum of a star. Using these parameters to select a suitable prior sample allows the prior to inform how the power spectra of stars with comparable global asteroseismic parameters behave. The other input parameters, \Teff and \bprp, are used to help inform on variables such as $\Gamma$ and \eps. These mode parameters are not always so well constrained by the PDS, while they have a strong dependence on \Teff~\citep{White_2012}, allowing for improved fitting of the $\ell=2,0$ and $\ell=1$ modes.

\subsubsection{Quality Control}

\begin{figure}
    \centering
    \includegraphics[width=0.96\linewidth]{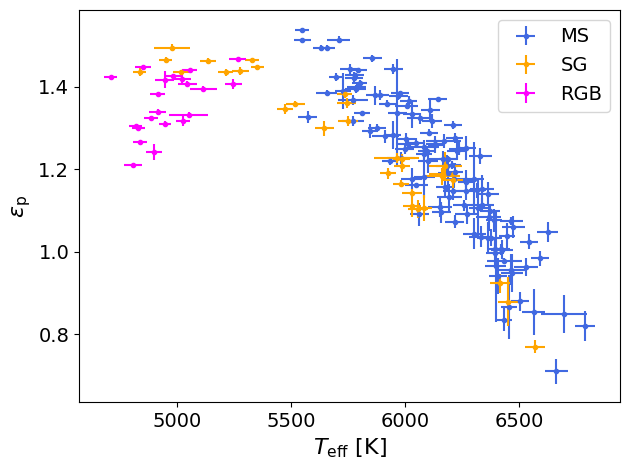}
    \caption{Relation between phase shift, \eps, and effective temperature, \Teff, for the stars in our sample.}
    \label{fig: Epsilon-Teff}
\end{figure}

To evaluate the quality of the mode identification we performed a manual inspection of the échelle diagram (where mode frequencies are shown as a function of frequency modulo \dnu;~\citealt*{Grec_1983}) and spectrum model for each of the stars. We also checked the values of \dnu, \numax used by \pbjam during mode identification to ensure that they corresponded with either the input parameters from~\citet{Lund_2024} or followed expected trends in the power spectrum. Lastly, we checked the asymptotic relation offset term \eps to ensure it aligned with the \Teff-\eps relation found by~\citet{White_2012} (see Fig.~\ref{fig: Epsilon-Teff}). We identified cases where \pbjam used the incorrect values for these asteroseismic parameters during mode identification, likely due to the star occupying an under-dense area in the prior parameters.

In the cases where \pbjam returned incorrect values of \dnu, \numax, or \eps, we manually defined prior distributions for the incorrect parameters to replace the automatically generated priors from \pbjam. We defined normally distributed priors, which for \dnu and \numax were centered on the estimate from~\citet{Lund_2024}, while for \eps, it was centered on an estimate from the relation from~\citet{White_2012}. The standard deviation for the normal distributions were 0.05-0.1\% for \dnu and 1-4\% for \numax and \eps. In the majority of cases, this narrower prior allowed \pbjam to find the correct estimate for these parameters.

\subsection{Final Peakbagging} \label{sec: Peakbagging}

The second main stage of \pbjam is the detailed peakbagging. The mode properties from the identification stage are passed onto the final peakbagging to allow a more detailed fitting of the modes to the spectrum. The frequencies, heights, and widths of the individual modes are now left as free variables, where the results from the mode identification stage are used as priors. The Lorentzian profile model from equation~\ref{eq: model} is again used for this final peakbagging stage, though now with fewer constraints. This allows the properties of the mode frequencies to be adjusted compared to the mode identification, which allows for effects such as acoustic glitches~\citep{Mazumbar_2014} and rotational splitting~\citep[e.g.][]{Ballot_2006,Hall_2021}, which may produce differences in the frequencies of the individual modes. The peakbagging is performed for modes of all angular degrees in the same step, allowing for correlations between the $\ell=2,0$ and $\ell=1$ modes.

The priors on the frequencies of the individual modes are $\beta$ distributions centered on the results from the mode identification stage, with the $\beta$ distribution shape parameters given by $\alpha=\beta=5$. The width of the $\beta$ prior is set to the small separation between $\ell=2$ and $\ell=0$ modes, $\delta\nu_{02}$, as this prevents the $\ell=0$ and $\ell=2$ modes from switching orders in any of the realisations of the posterior samples. The priors on mode heights and widths are normal distributions centered on the posterior result from the mode identification stage, with a standard deviation of 50\%. There were a few low \numax stars in the sample that had too high posterior mode widths from the detailed peakbagging stage. For these stars, tighter $\beta$ priors were applied to the mode widths to prevent them from becoming too high. For this detailed peakbagging, the \texttt{emcee} python library~\citep{Foreman_Mackey_2013} was used to perform the sampling of the parameters, using a Markov chain Monte Carlo algorithm. For more details on the peakbagging process used in \pbjam, see~\citet{Nielsen_2021, Nielsen_2025}.

The posterior distribution for each mode is the product of the prior of the model and the likelihood of the data from the light curve given the model. To enable validation of the results we provide a comparison between the standard deviation of the $\beta$ prior distribution and the median absolute deviation (MAD) of the posterior distributions. When no new information is gained concerning the mode location, the sampler recovers the prior distribution, in which case this ratio is unity. On the other hand when new information is gained the posterior to prior width ratio is less than unity. We provide these width ratios for each mode frequency included in the model for each star, with a potential cut-off of the posterior being 2/3 of the prior to be used as a potential guideline for when the likelihood is informative. This ensures that the spectrum provides enough new information on the mode frequency.

\section{Peakbagging results} \label{sec: Results}

In this section, we provide the results from the peakbagging process. The results for the individual modes have not been corrected for the Doppler-shift caused by the radial velocities of the stars, though methods for doing so are discussed by~\citet{Davies_2014}. The results on the global asteroseismic parameters and individual mode properties are presented in Sec.~\ref{sec: Global Results}, and the results for a few example stars are shown in Sec.~\ref{sec: Example Results}.

\subsection{Global results} \label{sec: Global Results}

The quality control process removed five stars\footnote{EPICs 211405262, 212728118, 228775782, 229159911 and 248776784} from the sample, due to the S/N of the PSD being too low for us to be confident in the peakbagging results provided by \pbjam. This left a total of 168 stars in the resulting sample. Fig.~\ref{fig: Star Types} shows an HR diagram for the sample, including the five stars that were removed. Of the 168 stars, 116 were treated with the MS model, 31 with the SG model, and 21 with the RGB model. Appendix A contains results for global asteroseismic parameters for all stars in table~\ref{tab: Global Results} and asteroseismic and modelling parameters specific to the SG and RGB stars in table~\ref{tab: Giant Results}. Figure~\ref{fig: All Parameters} shows the relation between the parameters in tables~\ref{tab: Global Results} and~\ref{tab: Giant Results} with \dnu and \Teff.

\begin{figure}
    \centering
    \includegraphics[width=0.96\linewidth]{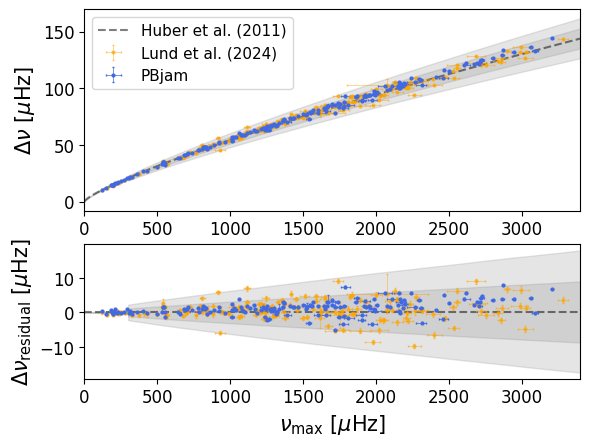}
    \caption{Relation between \numax and \dnu for the stars in our sample. The top cell shows the results from~\citet{Lund_2024} and \pbjam, with the dashed line being the empirical relation from~\citet{Huber_2011}. The shaded grey regions show the 1- and 2-$\sigma$ uncertainties on the empirical relation. The bottom cell shows the residuals of these resulting datasets against the empirical relation.}
    \label{fig: Delta dnu}
\end{figure}

\pbjam uses the asymptotic relation in equation~\ref{eq: asymptotic} to calculate estimates on \numax and \dnu, which allows for the curvature of the p modes with radial order. We compare these global asteroseismic parameters provided by \pbjam and those from~\citet{Lund_2024}. Fig.~\ref{fig: Delta dnu} shows this comparison, along with a comparison to the empirical scaling relation from~\citet{Huber_2011}. The results we report follow a similar trend to those from the CV pipeline used by~\citet{Lund_2024}, suggesting an accurate determination of these parameters. The precision on the individual stars is now improved with the \pbjam results, as well as the overall spread in the dataset being reduced.

The stellar parameters \numax, \dnu, and \Teff can be used in the scaling relations for the radius and mass of the star (e.g.~\citealt{Garc_a_2019})
\begin{equation}
    R\approx \text R_\odot \left(\frac{\dnu_\odot}{\dnu}\right)^2\left(\frac{\numax}{\nu_{\mathrm{max,\odot}}}\right)\left(\frac{\textit{T}_{\mathrm{eff}}}{\text T_{\mathrm{eff,\odot}}}\right)^{1/2},
    \label{eq: Radius}
\end{equation}

\begin{equation}
    M\approx \text M_\odot \left(\frac{\dnu_\odot}{\dnu}\right)^4\left(\frac{\numax}{\nu_{\mathrm{max,\odot}}}\right)^3\left(\frac{\textit{T}_{\mathrm{eff}}}{\text T_{\mathrm{eff,\odot}}}\right)^{3/2}.
    \label{eq: Mass}
\end{equation}
Using these scaling relations we can calculate the difference between the fractional uncertainties on the stellar radius and mass based on the estimates from~\citet{Lund_2024} and \pbjam.~\citet{Lund_2024} estimates provide median uncertainties of 2.5\% and 6.9\% on radius and mass, respectively, while the results from \pbjam provides median uncertainties of 1.3\% and 3.6\%. This is a reduction on the fractional uncertainties of 49\% on radius and 48\% on mass. We compared these radius results to those from \gaia DR3~\citep{Ulla_2022, Fouesneau_2023}. The standard deviation of the difference between the radii from~\citet{Lund_2024} and \pbjam compared to \gaia were 0.161 and 0.147, respectively.

\begin{figure*}
    \centering
    \includegraphics[width=0.495\linewidth]{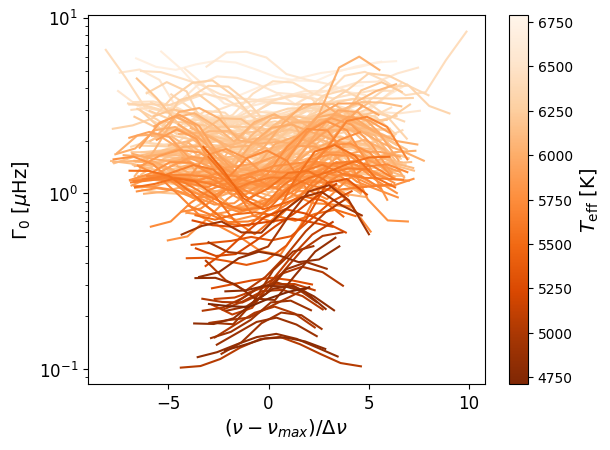}
    \includegraphics[width=0.49\linewidth]{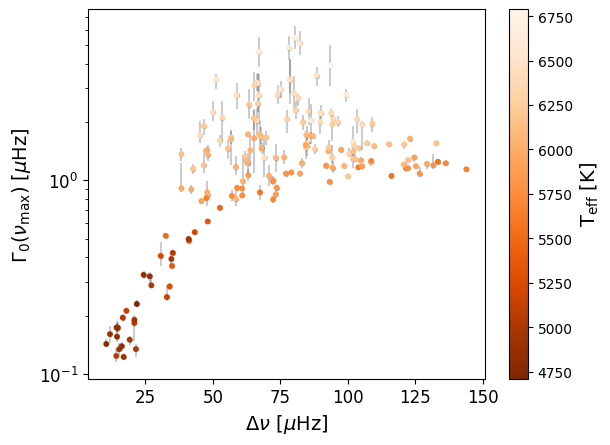}
    \caption{Mode width results for the radial modes. The left panel shows the radial mode widths against a proxy for radial orders, with each line representing the mode widths for a single star. The colour of the lines indicates the \Teff of the star. The data has had an Epanechnikov filter~\citep{Epanechnikov_1969,Lund_2017} applied to aid in the visualisation of the data. The right panel shows the radial mode width at \numax varying with \dnu.}
    \label{fig: Linewidths}
\end{figure*}

The mode width results provided by \pbjam are shown in Fig.~\ref{fig: Linewidths}. The right panel shows the how the mode width at \numax increases with \dnu as well as \Teff, as indicated by~\citet{Appourchaux_2014,Lund_2017}. This demonstrates the mode width dependency on both \Teff and mean density. The width of the modes for an individual star also increases with frequency, with a `dip' close to \numax. The dip in mode width below \numax decreases in amplitude with \Teff, so stars with a higher temperature have a more consistent mode width across the radial orders (see also~\citealt{Appourchaux_2014,Lund_2017}).

\begin{figure*}
    \centering
    \includegraphics[width=0.49\linewidth]{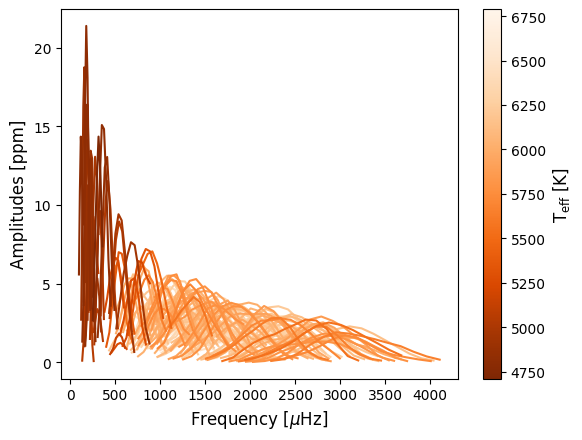}
    \includegraphics[width=0.495\linewidth]{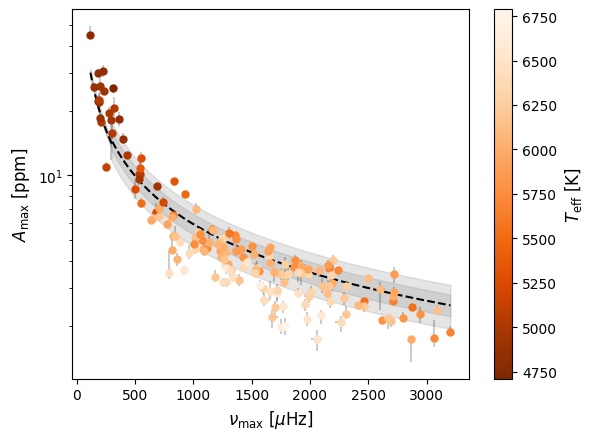}
    \caption{Amplitude results for the radial modes and oscillation envelope. The left cell shows the amplitude envelope of the radial modes, with an Epanechnikov filter applied to aid in the visualisation of the results. The colour scale shows the relation with \Teff. The right cell shows the maximum amplitude of the oscillation envelope for each of the stars. The dashed line shows the empirical relations from~\citet{Huber_2011} for a star with \Teff of 6000K. The shaded grey regions show the 1- and 2-$\sigma$ uncertainties on the empirical relation, where below $300\mu$Hz the~\citet{Huber_2011} relation has reduced uncertainties.}
    \label{fig: Amplitudes}
\end{figure*}

The mode heights measured by \pbjam were converted to mode amplitudes~\citep{Chaplin_2009}, and are shown in Fig.~\ref{fig: Amplitudes}. The increasing amplitude with decreasing \numax is a trend expected from previous studies (e.g.~\citealt{Kjeldsen_2005}). The right panel of Fig.~\ref{fig: Amplitudes} shows how the amplitude of the envelope varies with numax. We found, a larger spread towards lower amplitudes for a given \numax than expected from empirical relations~\citep{Huber_2011,Corsaro_2013}. This could be explained by the presence of high stellar activity in many of the stars in the sample dampening the amplitude of the modes~\citep[][see also~\citealt{Campante_2014,Sayeed_2025} for scaling relations with activity dependencies]{Chaplin_2011, Mathur_2019,Bessila_2024}. The scaling relation used does not account for the higher \Teff stars approaching the $\delta$-Scuti instability strip~\citep{Houdek_1999}, which could cause a dampening of the mode amplitudes~\citep{Chaplin_2011b}. This could provide another reason for the increased scatter from the scaling relation for the stars with higher \Teff values.

\begin{figure}
    \centering
    \includegraphics[width=0.96\linewidth]{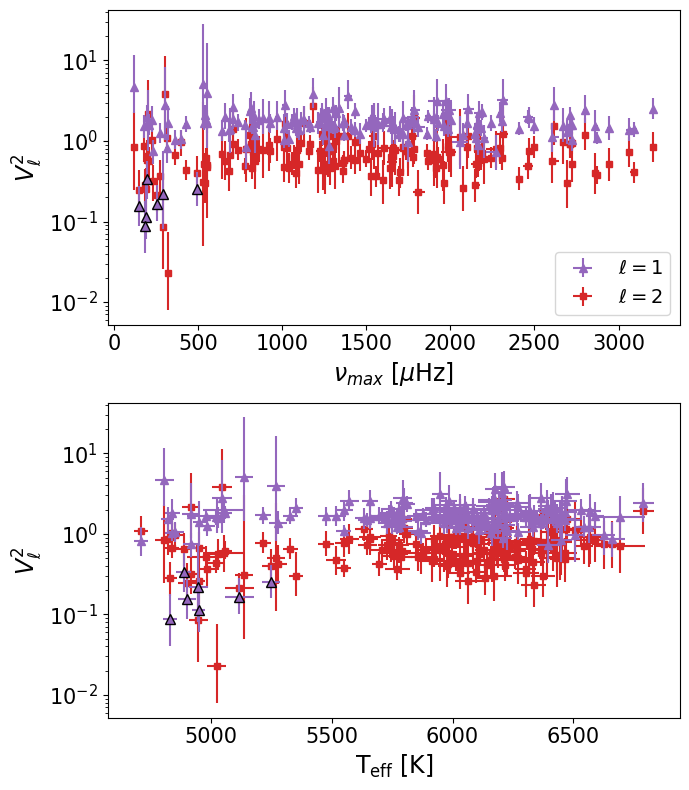}
    \caption{Mode visibilities for the $\ell=1$ modes (purple triangles) and $\ell=2$ modes (red squares), relative to the radial modes, shown against \numax in the top cell and against \Teff in the bottom cell. The seven $\ell=1$ results highlighted with a black outline show potential dipole-suppressed stars.}
    \label{fig: Visibilities}
\end{figure}

The amplitude results for the $\ell=1$ and $\ell=2$ modes can be compared to those of the $\ell=0$ modes to estimate the mode visibilities, following~\citep{Mosser_2012},
\begin{equation}
    V_\ell^2 = \frac{\langle A_\ell^2\rangle}{\langle A_0^2\rangle}.
    \label{eq: Visibilities}
\end{equation}
Fig.~\ref{fig: Visibilities} shows the results, with the median visibility for each angular degree being $V_1^2=1.65\pm0.45$ and $V_2^2=0.68\pm0.27$, where the uncertainties are defined as the standardised median-absolute-deviation. These results are higher than results from~\citet{Lund_2017}, where they found $V_1^2\approx1.50$ and $V_2^2\approx0.62$, and theoretical estimates from~\citet*{Ballot_2011} for stars in this range of \Teff, where $V_1^2\approx1.51$ and $V_2^2\approx0.53$. The visibilities resulting from the mode heights from \pbjam have large uncertainties which encompass the expected results from previous work. The resulting large spread in visibilities could be an explanation for these discrepancies.

Fig.~\ref{fig: Visibilities} shows seven stars highlighted with a black outline (EPICs 211411553, 229110376, 246154489, 246184564, 248696061, 249175306, 249639834) with dipole visibilities below the threshold used by~\citet{Stello_2016} defining stars of normal dipole modes and those with suppressed dipoles~\citep{Mosser_2012,Garc_a_2014}.~\citet{Fuller_2015,Stello_2016} have suggested that this reduction in dipole mode visibility is the result of strong magnetic fields. The \numax of these seven stars range from 180$\mu$Hz to 500$\mu$Hz, which is higher than many of the previously identified dipole-suppressed stars~\citep{Fuller_2015,Stello_2016,Lin_2025}. There are also stars with $\ell=2$ mode visibilities far lower than the median visibility, suggesting suppressed quadrupole modes.

\subsection{Example Results} \label{sec: Example Results}

\begin{figure*}
    \centering
    \includegraphics[width=0.38\linewidth]{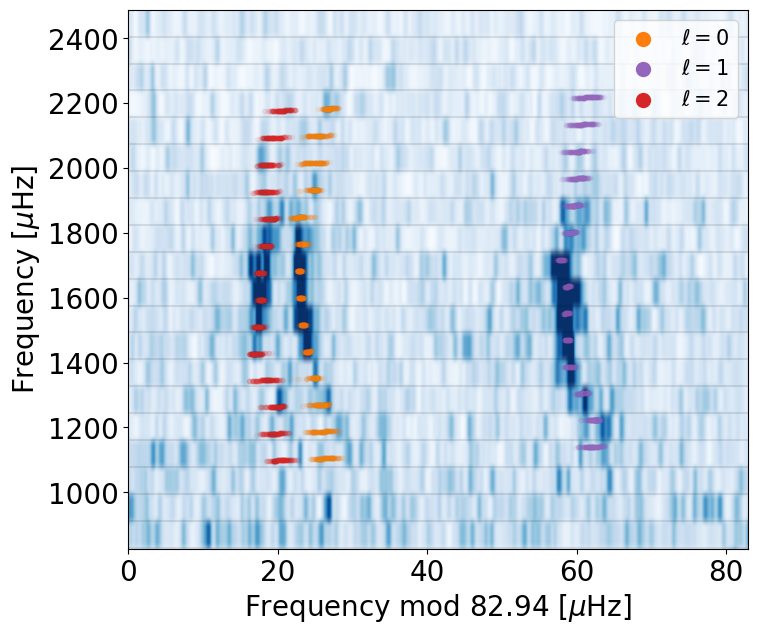}
    \includegraphics[width=0.58\linewidth]{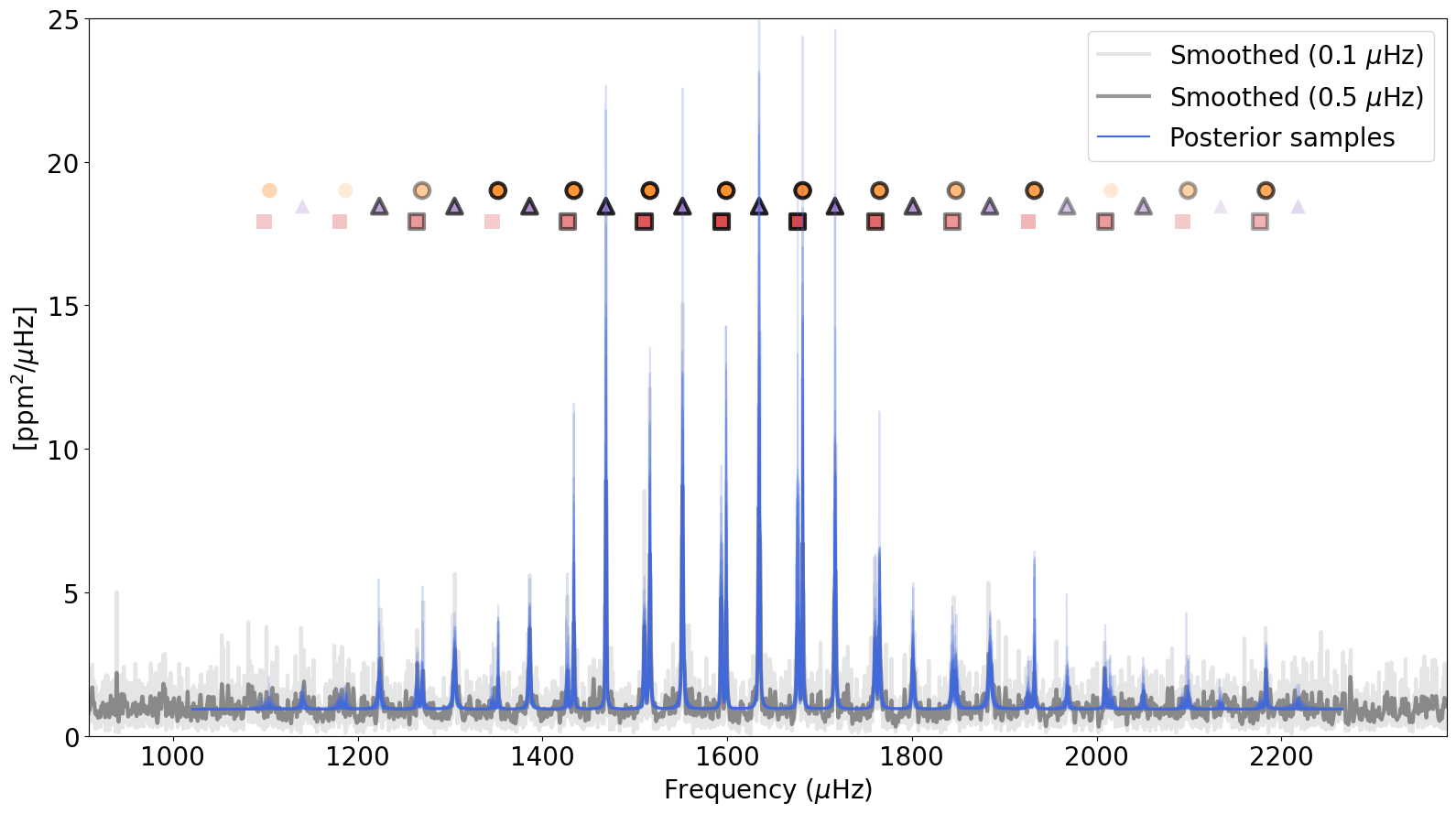}
    \caption{The final peakbagged results for EPIC 250165973. The left frame shows an échelle diagram, with the identified $\ell=2,0,1$ modes lying along the appropriate ridges of high power density. The right frame shows the spectrum, with two levels of smoothing of the data in grey and the peakbagged results overlayed in blue. The symbols representing the $\ell=0$ (orange circles), $\ell=1$ (purple triangles) and $\ell=2$ (red squares) modes show the location of the different angular degrees within the power spectrum. The opacity of the symbols represents the precision on the frequency of the mode, with higher opacity showing higher precision. The modes outlined in bold have validation results that passed the suggested threshold}.
    \label{fig: Gallifrey}
\end{figure*}

\begin{figure*}
    \centering
    \includegraphics[width=0.38\linewidth]{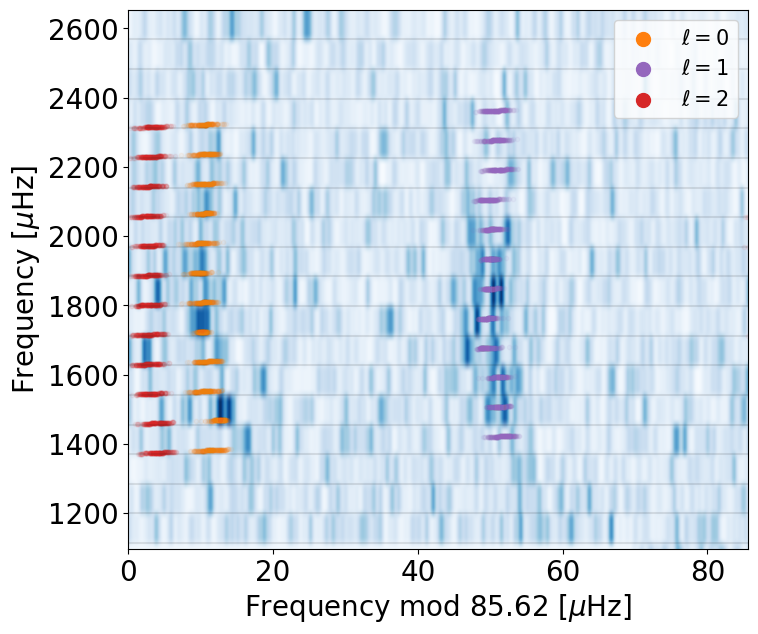}
    \includegraphics[width=0.58\linewidth]{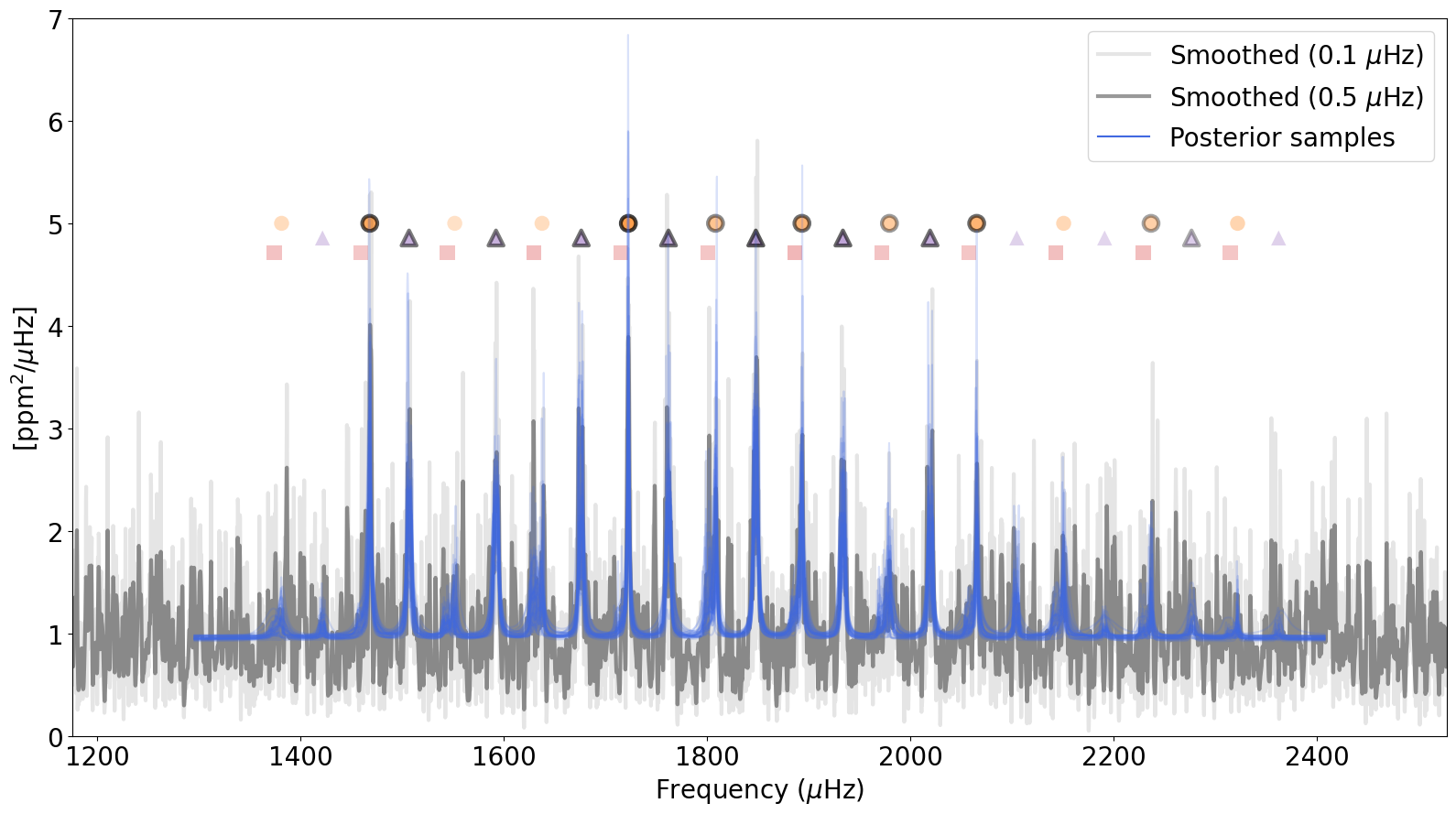}
    \caption{The final peakbagged results for EPIC 211401787. The échelle diagram shows higher levels of background noise, and lower precision on the identified modes, resulting in fewer radial orders being fitted. The right frame shows the posterior samples not reaching power densities too much higher than the background noise level.}
    \label{fig: Patrick}
\end{figure*}

\begin{figure*}
    \centering
    \includegraphics[width=0.38\linewidth]{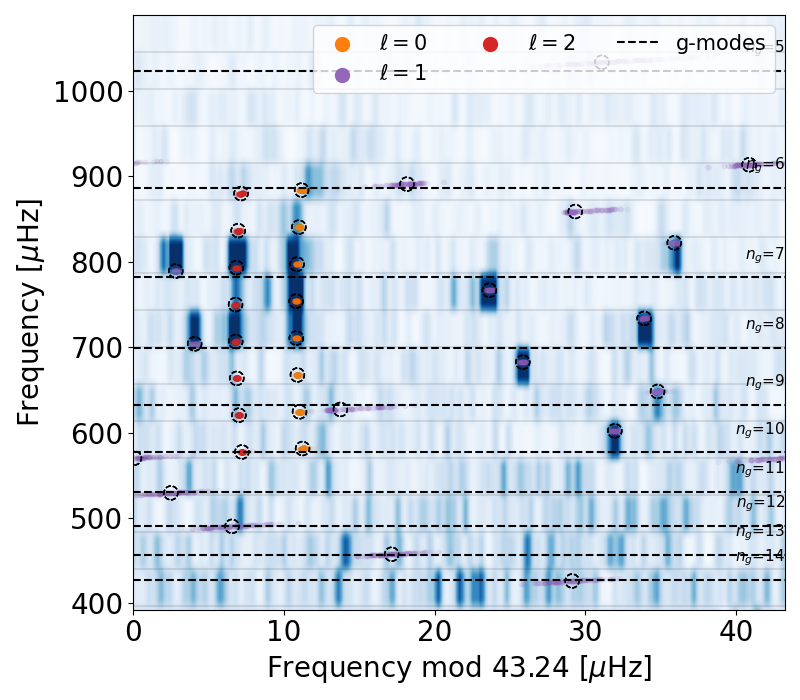}
    \includegraphics[width=0.58\linewidth]{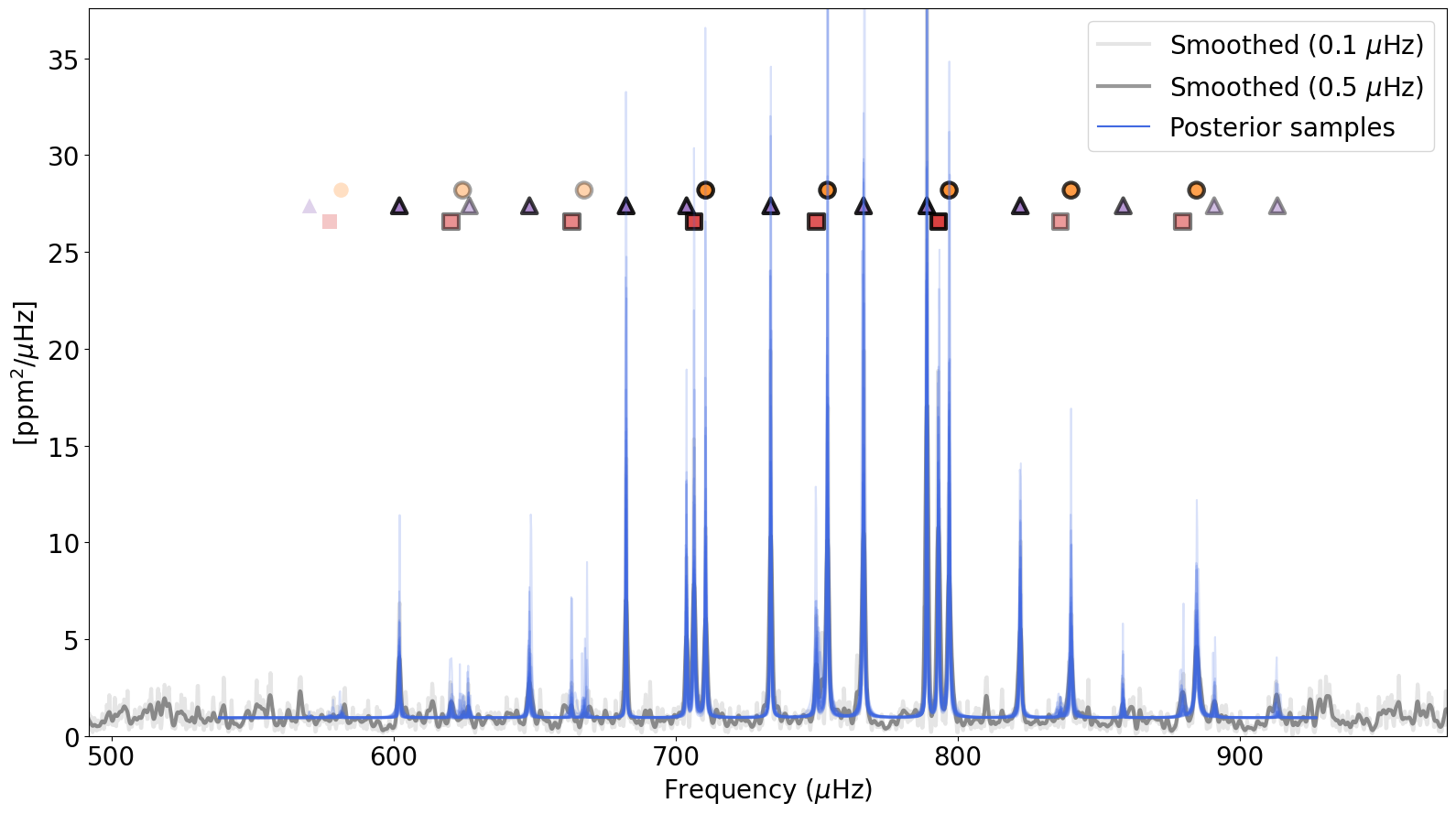}
    \caption{The results for EPIC 236224056, with a sub-giant model being used for the fitting of this star. The échelle diagram shows the results from the mode identification stage of \pbjam. The mixed $\ell$=1 modes can be seen in the diagram with the modes identified over top. The array of g-modes can also be seen across the p-mode envelope.}
    \label{fig: Father Christmas}
\end{figure*}

Figures~\ref{fig: Gallifrey},~\ref{fig: Patrick}, and~\ref{fig: Father Christmas} show examples of high S/N MS, low S/N MS and SG stars, respectively. The high S/N case shows EPIC 250165973. The high levels of S/N allowed for the fitting of 14 radial orders with 31 individual mode frequencies passing through our validation guideline. The clear signals from this star also allowed for the features such as rotational splitting and acoustic glitches to be visible. The low S/N case shows EPIC 211401787, for which 12 radial orders were fitted, with 15 individual modes passing through our validation guide. For EPIC 236224056, the sub-giant model was used to fit the dipole modes, since the mixed modes in this star are clearly visible. Table~\ref{tab: Individual Modes} provides results for mode parameters for all stars, as well as the posterior-prior width ratio validation results.

\begin{table*}
    \caption{Mode property and validation results. Only a few lines of the table are shown as an example. Full table of results for all characterised modes for all 168 stars available in an online table. Radial orders for $\ell=1$ modes for SG and RGB stars listed as $n=-1$.}
    \begin{tabular}{ccccccc}
    \hline
    EPIC & l & n & Frequency ($\mu$Hz) & Amplitude (ppm) & Mode width  ($\mu$Hz) & Posterior-prior width ratio\\
     
    \hline
    \\[\dimexpr-\normalbaselineskip+2pt]

    250165973  & 1 & 13 & 1223.62 $_{-0.61}^{+\,0.47}$ & 2.90 $_{-0.57}^{+\,0.54}$ & 2.64 $_{-0.99}^{+\,1.47}$ & 0.40 \\[3pt]
    250165973  & 1 & 14 & 1304.93 $_{-0.41}^{+\,0.42}$ & 3.37 $_{-0.49}^{+\,0.51}$ & 2.01 $_{-0.70}^{+\,0.96}$ & 0.31 \\[3pt]
    250165973  & 1 & 15 & 1386.16 $_{-0.33}^{+\,0.32}$ & 3.62 $_{-0.50}^{+\,0.54}$ & 1.57 $_{-0.59}^{+\,0.84}$ & 0.25 \\[3pt]
    250165973  & 0 & 13 & 1187.01 $_{-1.08}^{+\,1.05}$ & 0.83 $_{-0.45}^{+\,0.87}$ & 1.35 $_{-0.97}^{+\,3.49}$ & 0.83 \\[3pt]

    \hline
    \end{tabular}
    \label{tab: Individual Modes}
\end{table*}

\section{Conclusion} \label{sec: Discussion}

The results from this study have provided mode characterisation for over 6000 individual mode frequencies for 168 solar-like oscillators. These individual modes were identified and peakbagged using \pbjam, which utilises a large prior dataset to ensure the properties of these modes aligns with previous observations. In the process of peakbagging, we have also derived new global asteroseismic parameters \numax and \dnu for each of the stars with improved precision on these parameters and a more consistent dataset as a whole. This resulted in an improved precision on calculations from scaling relations on radius and mass of $49\%$ and $48\%$, respectively, compared to previous estimates of \numax and \dnu from~\citet{Lund_2024}.

The mode width results we obtained from \pbjam followed expected trends from previous studies and theoretical models, with the mode width at \numax increasing with \Teff and the amplitude of the drop in mode width at frequencies just below \numax decreasing with \Teff. The mode amplitude results show the expected trend based on scaling relations from~\citet{Huber_2011}. The median mode visibilites of the $\ell=1$ and $\ell=2$ modes were higher than expected, suggesting a potential underestimation in the amplitude of the radial modes. The mode visibility results suggested seven red giant stars possessed potentially suppressed dipole modes.

Approximately $25\%$ of the stars in the sample required manual priors on \dnu, \numax, or \eps to be applied for the mode identification due to the default prior sampling achieving incorrect estimates for these parameters. A likely common cause of this was an under-density in the large prior dataset for these particular stars. The results from this study can be added to the \pbjam prior and improve the prior sampling for stars of similar properties and thus lessen the occurrence rate of the prior finding incorrect estimates for these parameters.

These new individual mode frequency results can also be used to perform detailed stellar modelling to improve the constraints on stellar properties for the stars in this sample. This will provide an enhanced understanding on the characteristics and evolution of not only these stars but also other similar stars. The future \plato mission~\citep{Rauer_2025} will provide many observations of solar-like oscillators. This analysis shows the potential of using the peakbagging algorithm of \pbjam to identify and characterise the individual mode frequencies, thereby increasing the ability to utilise the observations from \plato.

\section*{Acknowledgments}

GTH, MBN, GRD, VS, and AS acknowledge support from the European Research Council (ERC) under the European Union’s Horizon 2020 research and innovation programme (CartographY G.A. n. 804752).
MBN and GRD acknowledge the support from the UK Space Agency.
MNL acknowledges support from the ESA PRODEX programme (PEA 4000142995).
RAG acknowledges the support from the GOLF and PLATO Centre National D'{\'{E}}tudes Spatiales grants.
SM acknowledges support by the Spanish Ministry of Science and Innovation with the grants number PID2019-107061GB-C66 and PID2023-149439NB-C41, and through AEI under the Severo Ochoa Centres of Excellence Programme 2020--2023 (CEX2019-000920-S).
The computations described in this paper were performed using the University of Birmingham's BlueBEAR HPC service, which provides a High Performance Computing service to the University's research community. See \url{http://www.birmingham.ac.uk/bear} for more details.
The authors acknowledge the dedicated teams behind the Kepler and K2 missions, without whom this work would not have been possible.
This work has made use of data from the European Space Agency (ESA) mission {\it Gaia} (\url{https://www.cosmos.esa.int/gaia}), processed by the {\it Gaia} Data Processing and Analysis Consortium (DPAC, \url{https://www.cosmos.esa.int/web/gaia/dpac/consortium}). Funding for the DPAC has been provided by national institutions, in particular the institutions participating in the {\it Gaia} Multilateral Agreement.

\textit{Software:} 
\texttt{PBjam} (\citealt{Nielsen_2021,Nielsen_2025},~\url{https://github.com/grd349/PBjam}),
\texttt{Python} \citep{python1995},
\texttt{Matplotlib} \citep{hunter2007},
\texttt{NumPy} \citep{Harris2020},
\texttt{pandas} \citep{Reback_2020},
\texttt{SciPy} \citep{Virtanen2020},
\texttt{Astropy} \citep{astropy2018},
\texttt{dynesty} \citep{Speagle_2020},
\texttt{emcee} \citep{Foreman_Mackey_2013},
\texttt{JAX} \citep{jax2018github},
\texttt{Echelle} (\url{https://github.com/danhey/echelle})

% Please add UKSA and CartographY acknowledgements :)

% Very well - Martin can give you the details

\section*{Data Availability}

Tables 1, A1 and A2 are available in online supplementary material. Other intermediate data products will be made available upon reasonable request.



\bibliographystyle{mnras}
\bibliography{ref} % if your bibtex file is called example.bib

%\begin{thebibliography}{99}
%\end{thebibliography}

%%%%%%%%%%%%%%%%%%%%%%%%%%%%%%%%%%%%%%%%%%%%%%%%%%

%%%%%%%%%%%%%%%%% APPENDICES %%%%%%%%%%%%%%%%%%%%%

\appendix

\section{Global asteroseismic results}

\begin{table*}
\centering
\caption{Results for global asteroseismic parameters from the mode identification. Only a few lines of the table are shown as an example. Full table of results for all 168 stars available in an online table.}
\begin{tabular}{cccccccc}
\hline\\[\dimexpr-\normalbaselineskip+1pt]
EPIC & \dnu ($\mu$Hz) & \numax ($\mu$Hz) & $\alpha_{\text p}$ ($10^{-3}$) & $\delta\nu_{01}$ ($\mu$Hz) & $\delta\nu_{02}$ ($\mu$Hz) & \eps & Nickname \\
\hline\\[\dimexpr-\normalbaselineskip+2pt]
201623069 & 58.47 $_{-0.08}^{+\,0.17}$ & 1083.28 $_{-5.06}^{+\,4.39}$ & 0.88 $_{-0.24}^{+\,0.44}$ & 27.34 $_{-0.42}^{+\,0.30}$ & 4.28 $_{-0.16}^{+\,0.16}$ & 1.11 $_{-0.05}^{+\,0.03}$ & Thing 1 \\\\[\dimexpr-\normalbaselineskip+2pt]
212487676 & 76.41 $_{-0.04}^{+\,0.05}$ & 1517.83 $_{-6.86}^{+\,4.77}$ & 1.53 $_{-0.22}^{+\,0.39}$ & 32.47 $_{-0.10}^{+\,0.09}$ & 6.20 $_{-0.34}^{+\,0.22}$ & 1.26 $_{-0.01}^{+\,0.01}$ & Aragorn \\\\[\dimexpr-\normalbaselineskip+2pt]
226083290 & 102.58 $_{-0.04}^{+\,0.09}$ & 2312.14 $_{-35.96}^{+\,33.21}$ & 0.93 $_{-0.08}^{+\,0.12}$ & 48.65 $_{-0.24}^{+\,0.23}$ & 6.35 $_{-0.21}^{+\,0.30}$ & 1.21 $_{-0.02}^{+\,0.01}$ & The Other One \\\\[\dimexpr-\normalbaselineskip+2pt]
228789925 & 60.83 $_{-0.03}^{+\,0.03}$ & 1114.27 $_{-2.69}^{+\,2.62}$ & 2.87 $_{-0.41}^{+\,0.28}$ & 29.50 $_{-0.60}^{+\,0.11}$ & 5.04 $_{-0.06}^{+\,0.06}$ & 1.38 $_{-0.01}^{+\,0.01}$ & Lightning McQueen \\\\[\dimexpr-\normalbaselineskip+2pt]
237799552 & 122.32 $_{-0.04}^{+\,0.04}$ & 2615.32 $_{-4.74}^{+\,5.65}$ & 1.04 $_{-0.08}^{+\,0.03}$ & 58.05 $_{-0.16}^{+\,0.17}$ & 7.99 $_{-0.05}^{+\,0.06}$ & 1.44 $_{-0.01}^{+\,0.01}$ & Zini \\\\[\dimexpr-\normalbaselineskip+2pt]
241011563 & 73.27 $_{-0.04}^{+\,0.10}$ & 1410.92 $_{-5.38}^{+\,5.08}$ & 1.26 $_{-0.07}^{+\,0.13}$ & 31.67 $_{-0.12}^{+\,0.08}$ & 5.32 $_{-0.08}^{+\,0.10}$ & 1.38 $_{-0.03}^{+\,0.01}$ & Thunder Star \\\\[\dimexpr-\normalbaselineskip+2pt]
246184564 & 11.84 $_{-0.02}^{+\,0.02}$ & 152.06 $_{-1.25}^{+\,2.27}$ & 5.25 $_{-0.39}^{+\,0.62}$ & 5.80 $_{-0.14}^{+\,0.33}$ & 1.42 $_{-0.06}^{+\,0.05}$ & 1.24 $_{-0.02}^{+\,0.02}$ & Dalek Sec \\\\[\dimexpr-\normalbaselineskip+2pt]
246358740 & 58.91 $_{-0.05}^{+\,0.07}$ & 1068.48 $_{-8.90}^{+\,6.49}$ & 0.67 $_{-0.17}^{+\,1.48}$ & 28.79 $_{-0.25}^{+\,0.32}$ & 5.10 $_{-0.15}^{+\,0.07}$ & 1.38 $_{-0.02}^{+\,0.01}$ & Gabbro \\\\[\dimexpr-\normalbaselineskip+2pt]
248665653 & 62.88 $_{-0.05}^{+\,0.03}$ & 1254.40 $_{-12.93}^{+\,13.48}$ & 0.75 $_{-0.01}^{+\,0.02}$ & 27.36 $_{-0.14}^{+\,0.15}$ & 4.17 $_{-0.19}^{+\,0.18}$ & 1.23 $_{-0.01}^{+\,0.01}$ & Bob the Builder \\\\[\dimexpr-\normalbaselineskip+2pt]
249151042 & 62.37 $_{-0.06}^{+\,0.05}$ & 1167.30 $_{-3.07}^{+\,2.48}$ & 1.17 $_{-0.26}^{+\,0.28}$ & 27.32 $_{-0.10}^{+\,0.10}$ & 5.16 $_{-0.10}^{+\,0.09}$ & 1.25 $_{-0.02}^{+\,0.02}$ & Bombur \\\\[\dimexpr-\normalbaselineskip+2pt]
\hline
\label{tab: Global Results}
\end{tabular}
\end{table*}

\begin{table*}
\centering
\caption{Results for asteroseismic parameters and mixed mode modelling parameters from the mode identification for the stars in the sample modelled treated with the SG and RGB models. Parameters $P_{\text L}$ and $P_{\text D}$ are mixing coefficients for the SG mixed mode model. Parameter $q$ is the coupling strength for the mixing of modes in the RGB model. Only a few lines of the table are shown as an example. Full table of results for all 52 SG and RGB stars available in an online table.}
\begin{tabular}{cccccccccccc}
\hline\\[\dimexpr-\normalbaselineskip+1pt]
EPIC & $\varepsilon_{\text g}$ & $\Delta\Pi_1$ (s) & $P_{\text L}$ ($10^{-3}$) & $P_{\text D}$ ($10^{-3}$) & q & Nickname \\
\hline\\[\dimexpr-\normalbaselineskip+2pt]
201623069 & 1.72 $_{-0.12}^{+\,0.08}$ & 424.74 $_{-62.00}^{+\,17.18}$ & 23.01 $_{-24.66}^{+\,16.07}$ & -1.12 $_{-3.18}^{+\,3.09}$ & .... & Thing 1 \\\\[\dimexpr-\normalbaselineskip+2pt]
211403248 & 1.52 $_{-0.03}^{+\,0.02}$ & 98.99 $_{-0.06}^{+\,0.08}$ & .... & .... & 0.12 $_{-0.01}^{+\,0.02}$ & Patty \\\\[\dimexpr-\normalbaselineskip+2pt]
211506655 & 1.47 $_{-0.02}^{+\,0.02}$ & 97.00 $_{-0.05}^{+\,0.07}$ & 10.69 $_{-1.13}^{+\,1.49}$ & 2.85 $_{-0.40}^{+\,0.34}$ & .... & Abi \\\\[\dimexpr-\normalbaselineskip+2pt]
211811597 & 0.74 $_{-0.05}^{+\,0.08}$ & 85.80 $_{-0.10}^{+\,0.06}$ & .... & .... & 0.15 $_{-0.01}^{+\,0.01}$ & Jocasta Nu \\\\[\dimexpr-\normalbaselineskip+2pt]
212586030 & 1.61 $_{-0.03}^{+\,0.04}$ & 95.67 $_{-0.12}^{+\,0.09}$ & .... & .... & 0.16 $_{-0.01}^{+\,0.01}$ & The Stranger \\\\[\dimexpr-\normalbaselineskip+2pt]
220222356 & 1.00 $_{-0.06}^{+\,0.03}$ & 85.67 $_{-0.05}^{+\,0.09}$ & .... & .... & 0.19 $_{-0.00}^{+\,0.01}$ & Darth Vader \\\\[\dimexpr-\normalbaselineskip+2pt]
228837389 & 1.74 $_{-0.01}^{+\,0.01}$ & 137.90 $_{-0.11}^{+\,0.10}$ & -18.27 $_{-0.30}^{+\,0.32}$ & -3.65 $_{-0.01}^{+\,0.01}$ & .... & Hatchling \\\\[\dimexpr-\normalbaselineskip+2pt]
236224056 & 1.46 $_{-0.05}^{+\,0.04}$ & 151.28 $_{-0.73}^{+\,0.87}$ & 25.33 $_{-1.02}^{+\,1.39}$ & -3.68 $_{-0.16}^{+\,0.19}$ & .... & Father Christmas \\\\[\dimexpr-\normalbaselineskip+2pt]
248855475 & 1.57 $_{-0.01}^{+\,0.01}$ & 140.71 $_{-0.19}^{+\,0.19}$ & 19.65 $_{-0.89}^{+\,0.71}$ & 2.85 $_{-0.29}^{+\,0.22}$ & .... & Romana \\\\[\dimexpr-\normalbaselineskip+2pt]
249175306 & 0.94 $_{-0.23}^{+\,0.12}$ & 85.63 $_{-2.14}^{+\,0.93}$ & .... & .... & 0.07 $_{-0.04}^{+\,0.03}$ & Superintendent Chalmers \\\\[\dimexpr-\normalbaselineskip+2pt]
\hline
\label{tab: Giant Results}
\end{tabular}
\end{table*}

\begin{figure*}
    \centering
    \includegraphics[width=0.96\linewidth]{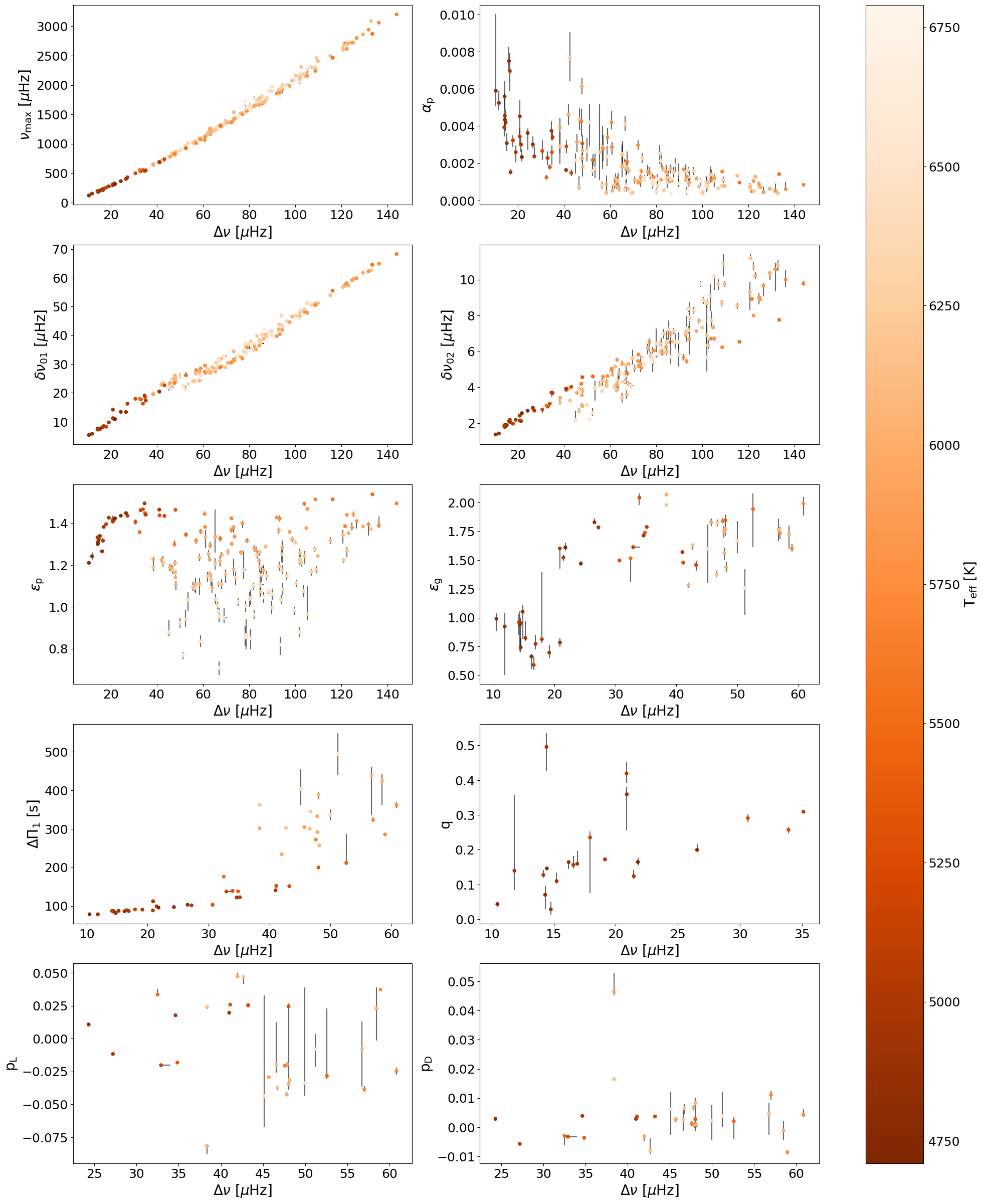}
    \caption{Global asteroseismic parameter results shown in tables~\ref{tab: Global Results} and~\ref{tab: Giant Results} against \dnu. The colour gradient represents the change in \Teff.}
    \label{fig: All Parameters}
\end{figure*}

%%%%%%%%%%%%%%%%%%%%%%%%%%%%%%%%%%%%%%%%%%%%%%%%%%

% Don't change these lines
\bsp	% typesetting comment
\label{lastpage}
\end{document}